\documentclass{elsart}
\usepackage{epsfig}
\usepackage{rotating}
\begin{document}

%TCIMACRO{
%\TeXButton{frontmatter}{\begin{frontmatter}
%\title{Radiation studies for GaAs in the ATLAS Inner Detector}
%\author{M.Rogalla, Th.Eich, N.Evans, R.Geppert, R.G\"oppert,}
%\author{R.Irsigler, J.Ludwig, K.Runge, Th.Schmid}
%\address{Albert-Ludwigs-Universit\"at Freiburg, Fakult\"at f\"ur Physik, D-79104 Freiburg im Breisgau}
%\begin{abstract}
%We estimate the hardness factors and the equivalent 1 MeV neutron fluences for hardrons fluences expected at the
%GaAs positions wheels in the ATLAS Inner Detector. On this basis the degradation of the GaAs particle detectors made from different 
%substrates as a function of years LHC operation is predicted.
%\end{abstract}
%\end{frontmatter}
%}}
%BeginExpansion
\begin{frontmatter}
\begin{flushright}
FREIBURG-EHEP-97-01
\end{flushright}
\date{25 June 1996}
\title{Radiation studies for GaAs in the ATLAS Inner Detector}
\author{M.Rogalla, Th.Eich, N.Evans, R.Geppert, R.G\"oppert,}
\author{R.Irsigler, J.Ludwig, K.Runge, Th.Schmid}
\address{Albert-Ludwigs-Universit\"at Freiburg, Fakult\"at f\"ur Physik, D-79104 Freiburg im Breisgau}
\begin{abstract}
We estimate the hardness factors and the equivalent 1 MeV neutron fluences 
for hadrons fluences expected at the
GaAs positions wheels in the ATLAS Inner Detector. On this basis the degradation of the GaAs particle detectors made from different 
substrates as a function of years LHC operation is predicted.
\end{abstract}
\end{frontmatter}
%
%EndExpansion

\section{Introduction}

It was proposed by the ATLAS collaboration to use GaAs strip detectors as a
part of the inner tracker at LHC because of the good radiation resistance to
fast neutron irradiation up to 4$\cdot $10$^{14}$ n cm$^{-2}$\cite{aachen}.
The high hadron radiation levels with fluences in the range of 10$^{13}$ cm$%
^{-2}$ yr$^{-1}$ in the forward region of the ATLAS inner detector will
effect the main parameters for the operation of the GaAs detectors: leakage
current density, space charge density, mean free drift length and charge
collection efficiency. In the first part of this paper we discuss the
radiation level for the GaAs wheels (R=26-32.7 cm,Z=85-170 cm)\cite{backup}
and calculate the hardness factors for the integrated neutron and proton
spectra. Then an attempt is made to estimate the 1 MeV neutron equivalent
fluence for a ten year LHC operation. In the second part we summarize damage
functions deduced experimentally from pad detectors irradiated in the ISIS
spallation neutron source at the Rutherford Appleton Laboratory\cite{edwards}
and with 24 GeV/c protons at the proton synchrotron (CERN).

\section{Radiation levels and extraction of equivalent 1 MeV neutron fluences
}

The radiation levels for the ATLAS inner detector have been calculated by
A.Ferrari\cite{ferrari} using the FLUKA code with DTUJET. Table 1 summarizes
the expected hadron radiation fluences assuming a p-p cross section of 80 mb
and a maximum luminosity of 1.0$\cdot $10$^{34}$ cm$^{-2}$ s$^{-1}$
(operating time 10$^7$ s per year) for the LHC. For the following we assume
that the charged hadron fluences only consists of pions and protons.

\begin{table}[tbph]
\caption{Mean, peak energy and fluences (10 years operation) of hadron
radiation level at the position of the GaAs wheels (R = 30 cm) in the ATLAS
inner detector.}
\begin{tabular}{|c|c|c|c|}
\hline
& peak energy [MeV] & mean energy [MeV] & fluence [cm$^{-2}$] \\ \hline
pions & 600 & 1900 & 1.86$\cdot $10$^{14}$ \\ \hline
charged hadrons & 840 & 230 & 3.05$\cdot $10$^{13}$ \\ 
without pions &  &  &  \\ \hline
neutrons & - & 28 & 1.54$\cdot $10$^{13}$ \\ \hline
\end{tabular}
\end{table}
The basic assumption to estimate the radiation damage for GaAs in the inner
detector is that all bulk damage effects scale linearly with the total
non-ionizing energy loss (NIEL) $D(E)$. This quantity has been calculated as
a function of kinetic energy for the interaction of several particles with
GaAs\cite{neutron}\cite{proton}. Using this functions it is possible to
relate the damage from monoenergetic 1 MeV neutrons to the damage from other
particles with different energies via the hardness factor $\kappa $. For a
radiation source with known energy spectrum $\varphi (E)$ and cut-offs at $%
E_{min}$ and $E_{max}$ we define 
\begin{equation}
\kappa =\frac 1{D(\mbox{1 MeV,n})}\frac{\int_{E_{\min }}^{E_{\max }}\varphi
(E)D(E)dE}{%
\begin{array}{c}
\int_{E_{\min }}^{E_{\max }}\varphi (E)dE
\end{array}
}
\end{equation}
Thus, for a measured fluence $\Phi _{meas}=\int_{E_{\min }}^{E_{\max
}}dE\varphi (E)$ ,the equivalent 1 MeV fluence is given by $\Phi =\kappa
\Phi _{meas}$. For the normalization factor $D($1MeV,n$)$ we use 3.1 keV/cm
calculated by E.A.Burke\cite{proton}. Figure 1 shows the estimated
non-ionizing energy loss of protons and neutrons as a function of kinetic
energy in GaAs. The solid lines are the calculated values of A.M. Ougouag%
\cite{neutron} and E.A.Burke\cite{proton} for protons and neutrons
respectively.

For neutrons there are only values up to 14 MeV available and we assume a
plateau of the NIEL for higher neutron energies (dashed line) as for Silicon
because of the expected decrease of the cross section and the saturation of
the Lindhard function. This uncertainty leads only to a small error of the
hardness factor for the neutron spectra in the inner detector because 98\%
of the neutrons have energies below 14 MeV. For protons the NIEL up to 1 GeV
is known. Additional we determine from the comparison of neutron and proton
irradiated detectors a hardness factor of 7 for 23 GeV protons\cite{NIM}. To
determine the hardness factors of the proton spectra in the inner detector
we extrapolate between 1 GeV and 23 GeV. At this time there is no
calculation of the NIEL for pions in GaAs. Only experimental data for the
radiation damage of 200 MeV pions are available\cite{aachen2} which give a
hardness factor of about 9 in comparison to proton irradiated samples. Due
to the delta resonance we expect a maximum of the NIEL as a function of
kinetic energy at 200 MeV (similar to Silicon). The peak and mean energy of
the pions expected at the GaAs wheel position is much higher and from this
we can estimate a value of about 8 as an upper limit of the hardness factor.
Table 2 summarizes the calculations of the hardness factors and also shows
the equivalent 1 MeV neutron fluence estimated for ten year LHC operation.
The total fluence in ten years comes up to a level of about 1.75$\cdot $10$%
^{15}$ cm$^{-2}$ and is mainly due to pions.

\begin{center}
\begin{table}[tp]
\caption{Hardness factors of the hardrons at R = 30 cm in the ATLAS inner
detector and estimate equivalent 1MeV neutron fluences (GaAs) for a ten year
LHC operation.}
\begin{tabular}{|c|c|c|}
\hline
& hardness factor $\kappa $ & $\Phi (1MeV,n)$[cm$^{-2}$] \\ \hline
pions & 8 & 1.48$\cdot $10$^{15}$ \\ \hline
protons & 8 & 2.5$\cdot $10$^{14}$ \\ \hline
neutrons & 0.5 & 7.7$\cdot $10$^{12}$ \\ \hline\hline
total &  & 1.75$\cdot $10$^{15}$ \\ \hline
\end{tabular}
\end{table}
\end{center}

\section{Detector parameters of GaAs as a function of LHC operation years}

The pad detectors, for investigation of radiation hardness, are Schottky
diodes made on semi-insulating GaAs with a thickness of 200 $\mu $m and a
diameter of 2 or 3 mm. The contacts and substrates are described elsewhere 
\cite{NIM}. The detectors were irradiated using neutrons (ISIS) and 23 GeV
protons (CERN). Figure 2 shows the leakage current density as a function of $%
\Phi ($1MeV,n$)$. We observe for high ohmic material a slight increase of
the leakage current density to a value of 30 nA/mm$^2$ at 20$^{\circ }$C and
for medium ohmic material a decrease down to the same value. This means the
leakage current density after ten year LHC operation is independent of the
value before irradiation and small in comparison to Si pad detectors \cite
{SI}. The typical behavior of the MIP signal height ($^{90}$Sr-Source) of
irradiated pad detectors is an exponential decrease down to about 4000 e$%
^{-} $ for an equivalent fluence of 1.3$^{.}$10$^{15}$ cm$^{-2}$ (e.g. wafer
MCP90 Figure 3). For the measurements we used a bias voltage of 200 V and a
shaping time of 500 ns. To investigate this behavior we determined the mean
free drift lengths by irradiating the detector from the front and backside
with alpha particles ($^{241}$Am). The charge collection efficiency (CCE) is
then given by 
\begin{equation}
CCE=\frac{\lambda _n}d\left[ 1-\exp \left( -\frac{d-x_0}{\lambda _n}\right)
\right] +\frac{\lambda _h}d\left[ 1-\exp \left( -\frac{x_0}{\lambda _h}%
\right) \right]
\end{equation}

where $\lambda _n$ and $\lambda _h$ denotes the drift lengths of electrons
and holes, $d$ the detector thickness and $x_0$ represents the generation
point of electron hole pairs. We assume that the mean free drift length is
independent of the position within the detector. In any case this is valid
for electrons because of the saturation of the drift velocity at electric
fields higher than 10$^4$ V/cm. For holes we observe a plateau of the CCE
for bias voltages higher than 200 V \cite{NIM} which indicates also a
saturation of the hole drift length. Figure 4 shows the mean free drift
length of electrons and holes (20 $^{\circ }$C, 300 V bias) as a function of
the 1 MeV equivalent neutron fluence for diodes from wafer FR41. We observe
a strong decrease of the mean free drift length for both carriers down to 32 
$\mu $m for electrons and 22 $\mu $m for holes. A comparison of different
materials shows that there is a dependence of the electron mean free drift
length on the resistivity. The material with the lowest resistivity shows
for all irradiation levels the highest electron mean free drift length but
is still going down to about 36 $\mu $m (Figure 5). CV-measurements have
shown that the space charge density decreases with increasing irradiation 
\cite{NIM}. Figure 6 plots shows corresponding bias voltage to make the
detector full active. This voltage decreases for diodes on Wafer FR76 from
250 V down to 10 V after ten years of LHC operation.

\section{Summary and Conclusion}

The calculation of the hardness factors of the proton and neutron spectra
and the estimation for pions result in a 1 MeV neutron equivalent fluence of
1.75$^{.}$10$^{15}$ cm$^{-2}$ at R = 30 cm in the ATLAS inner detector for
ten years of LHC operation. Therefore we expect at 20 $^{\circ }$C for the
GaAs detectors leakage current density of 30 nA/mm$^2$ independent of the
resistivity of the semi-insulating GaAs substrates. This value is low in
comparison to silicon detectors. Also the voltage to make the detector fully
active is smaller than 200 V. The signal loss due to trapping and the
reduction of the carrier mobility is a severe problem. But the signal height
of pad and strip detectors are not the same, because of the different
weighting fields. A concept proposed by Th.Schmid et al. \cite{thilo}
predicts for a strip detector with special bias a signal height of 7000 e$%
^{-}$ considering only the electrons with a mean free drift length of 40 $%
\mu $m after ten years of LHC operation.

\section{Acknowledgments}

We are grateful to M.Edwards (RAL) and F.Lemeilleur (CERN) for the help with
the irradiation. We also thank A.Ferrari for providing the data of the
radiation levels in the ATLAS inner detector. This work has been supported
by the BMBF under contract 057FR11I.

\newpage\

\begin{figure}[htbp]
   \begin{center}
%   \begin{turn}{-90}
      \mbox{
          \epsfxsize=13.50cm
           \epsffile{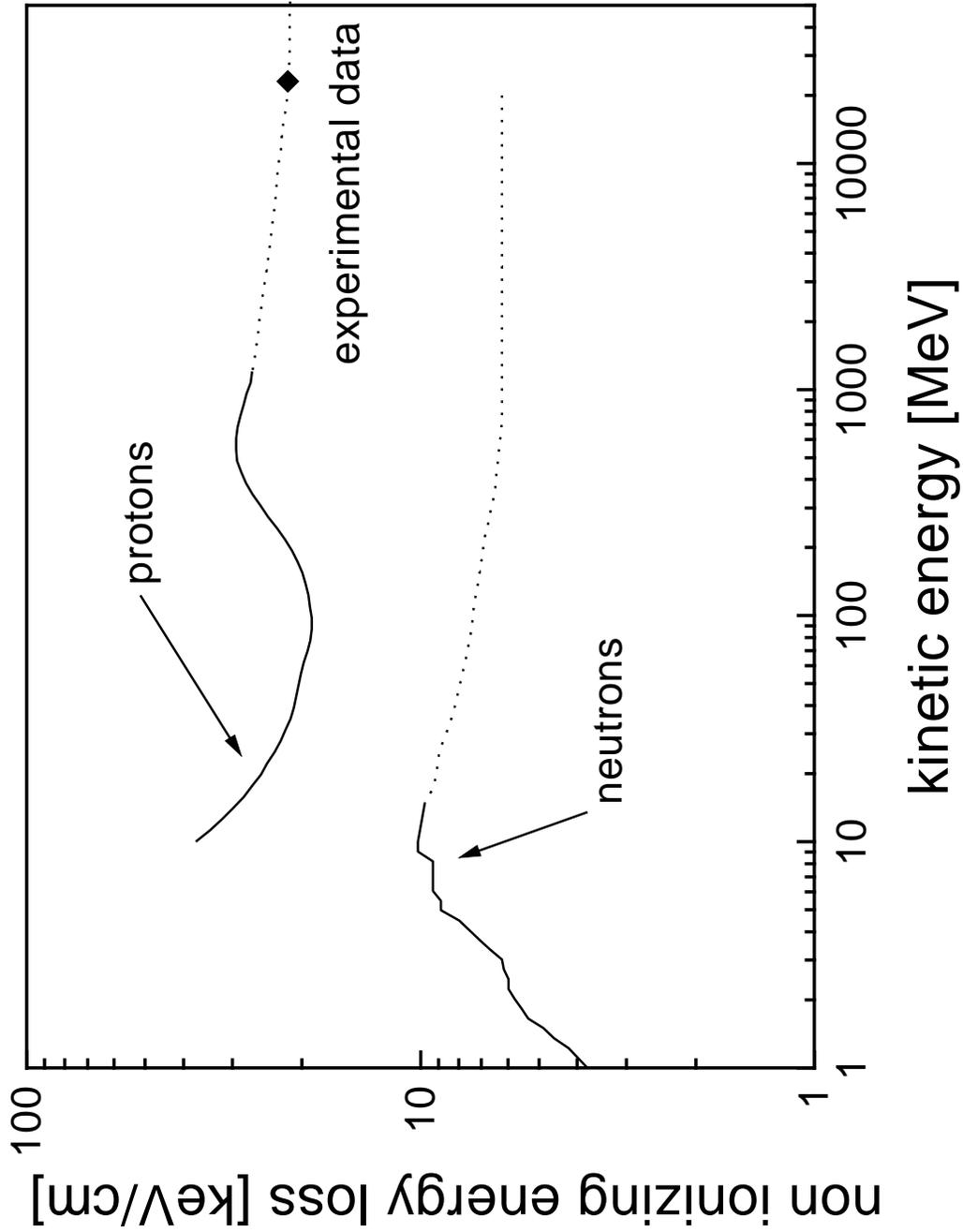}
           }
%\end{turn}
   \end{center}
\caption{
The estimated non ionizing energy loss (NIEL) of protons and
neutron as a function of kinetic energy in GaAs.
}
\end{figure}

\begin{figure}[htbp]
   \begin{center}
      \mbox{
          \epsfxsize=13.50cm
           \epsffile{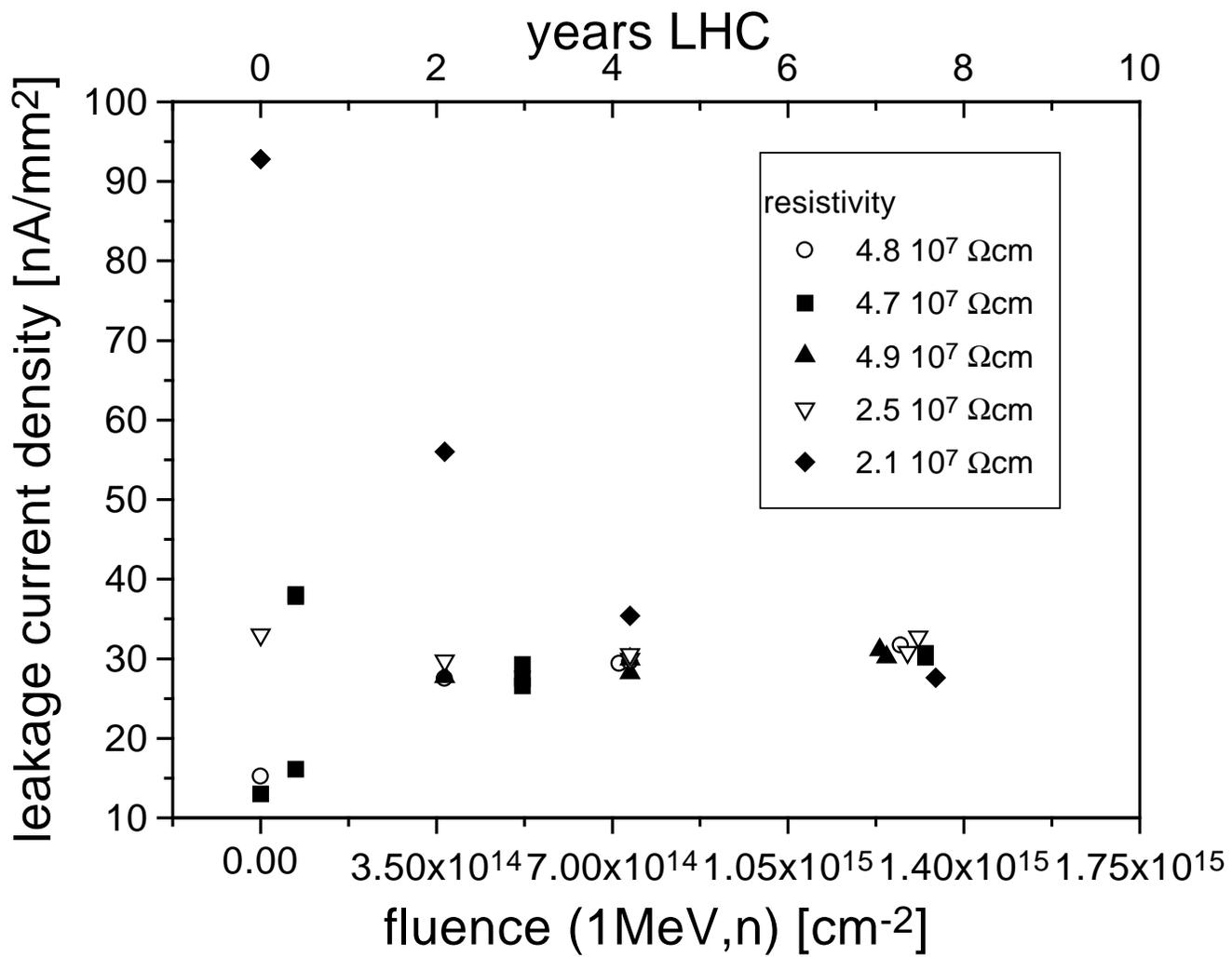}
           }
   \end{center}
\caption{
The leakage current densities at 20$^{\circ }$C of Schottky diodes
made on different substrates versus the 1 MeV neutron equivalent fluence.
}
\end{figure}

\begin{figure}[htbp]
   \begin{center}
          \epsfxsize=13.50cm
           \epsffile{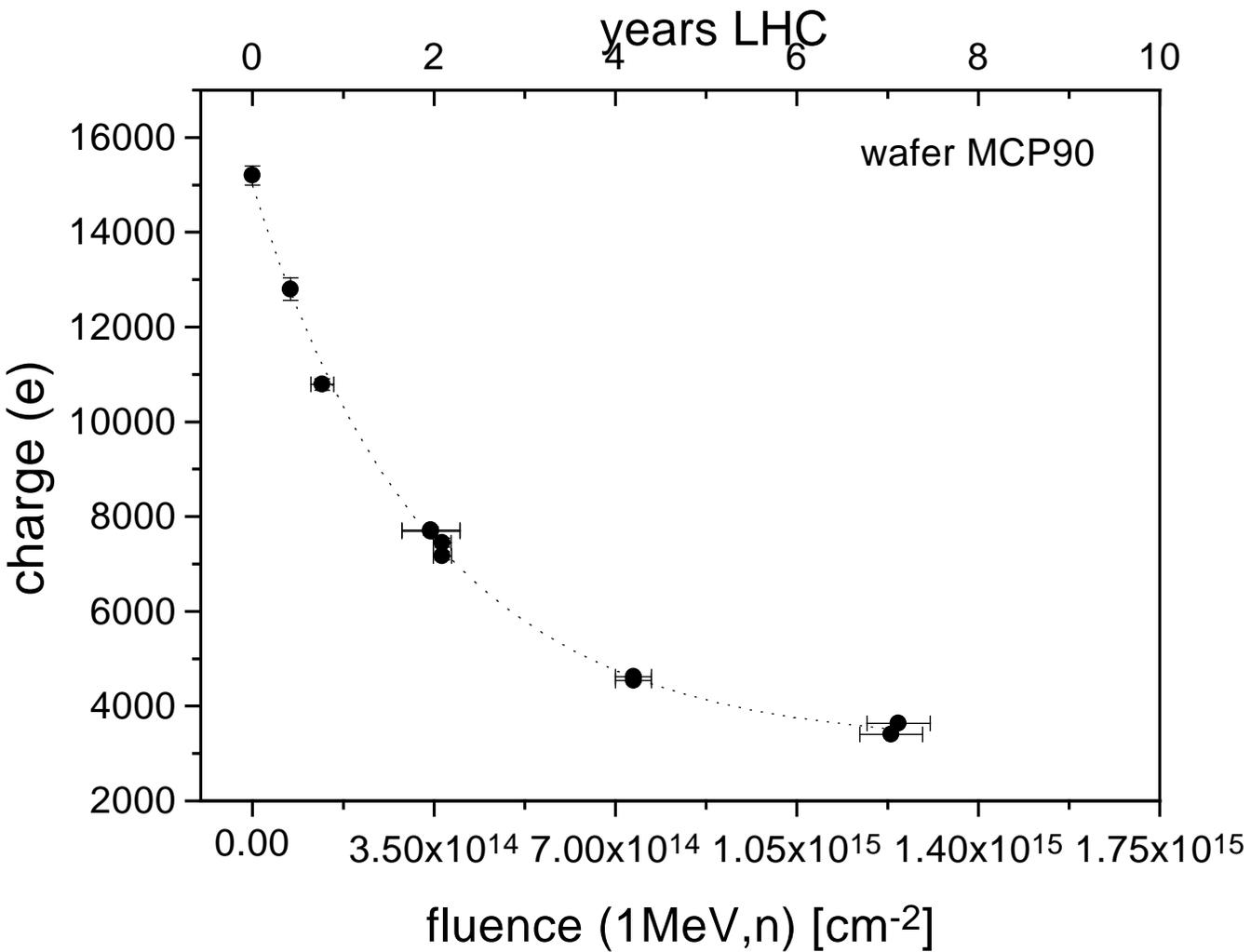}
   \end{center}
\caption{
The typical behavior of the MIP signal height as a function of the
1 MeV neutron equivalent fluence ($^{90}$Sr, 200 V bias, 500ns shaping time,
wafer MCP90).
}
\end{figure}

\begin{figure}[htbp]
   \begin{center}
      \mbox{
          \epsfxsize=13.50cm
           \epsffile{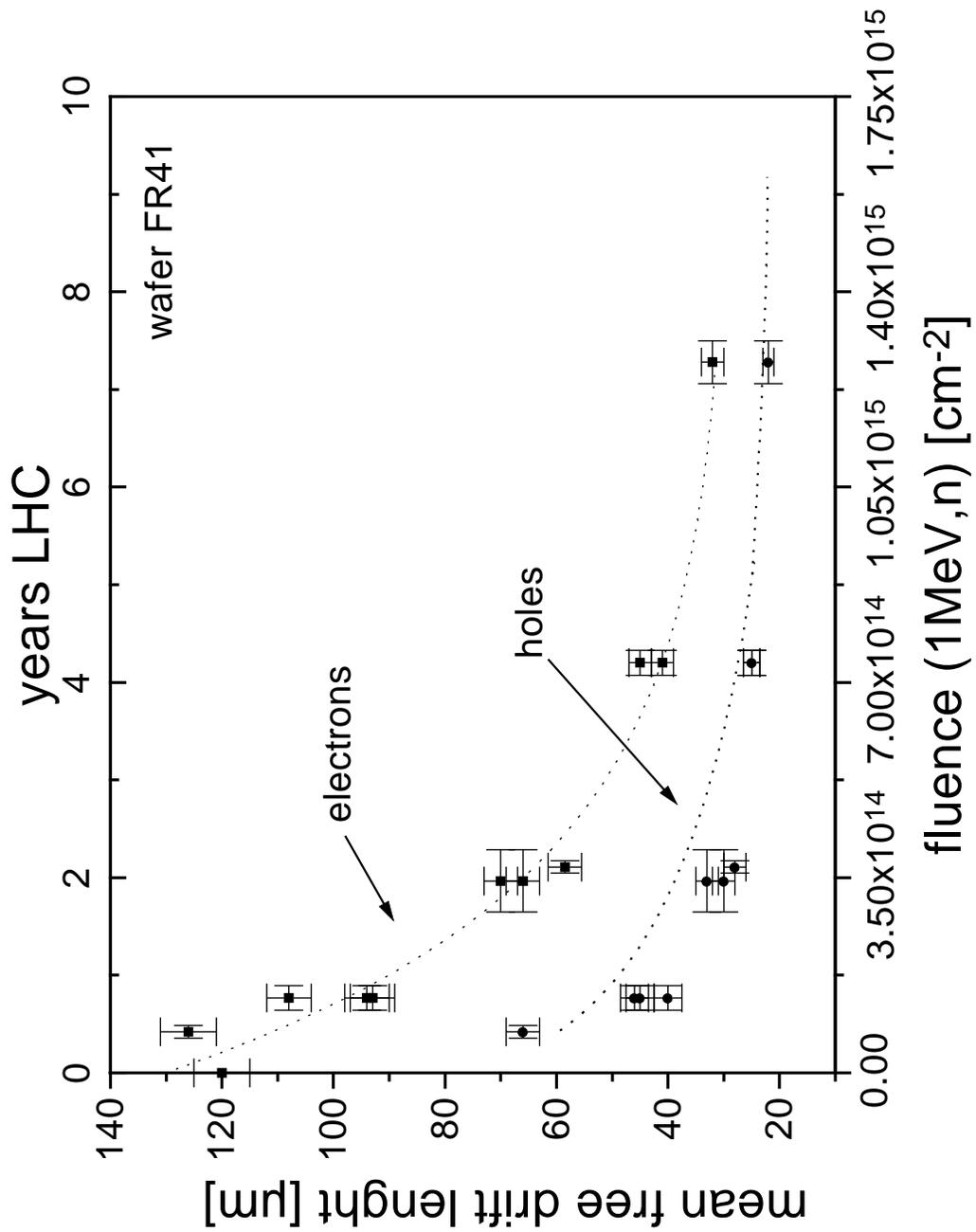}
           }
   \end{center}
\caption{
The mean free drift length of electrons and holes obtained from
alpha spectra as a function of the irradiation level (20$^{\circ }$C, 300 V
bias, wafer FR41).
}
\end{figure}

\begin{figure}[htbp]
   \begin{center}
      \mbox{
          \epsfxsize=13.50cm
           \epsffile{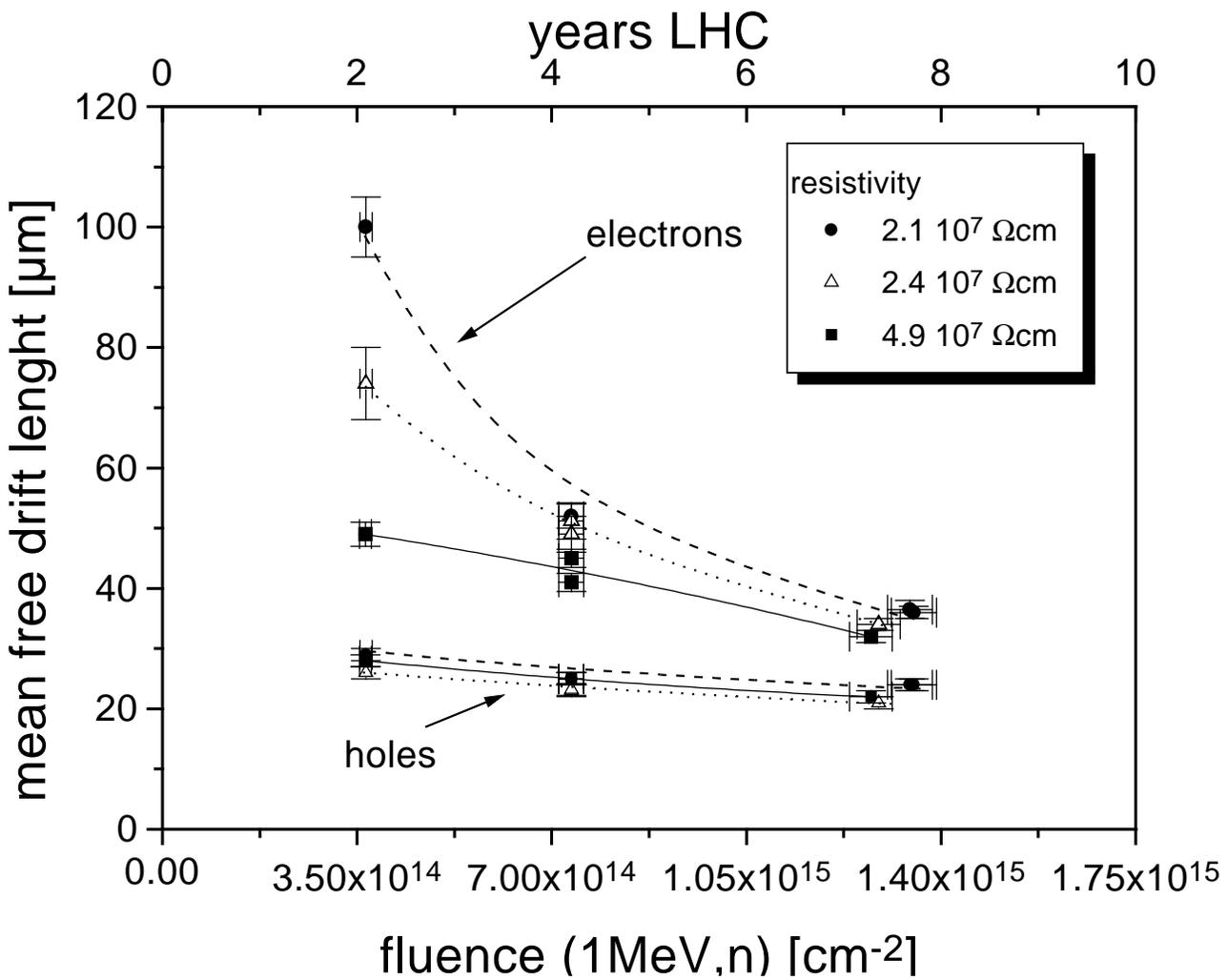}
           }
   \end{center}
\caption{
The mean free drift length of electrons and holes versus the 1 MeV
neutron equivalent fluence for different materials.
}
\end{figure}

\begin{figure}[htbp]
   \begin{center}
      \mbox{
          \epsfxsize=13.50cm
           \epsffile{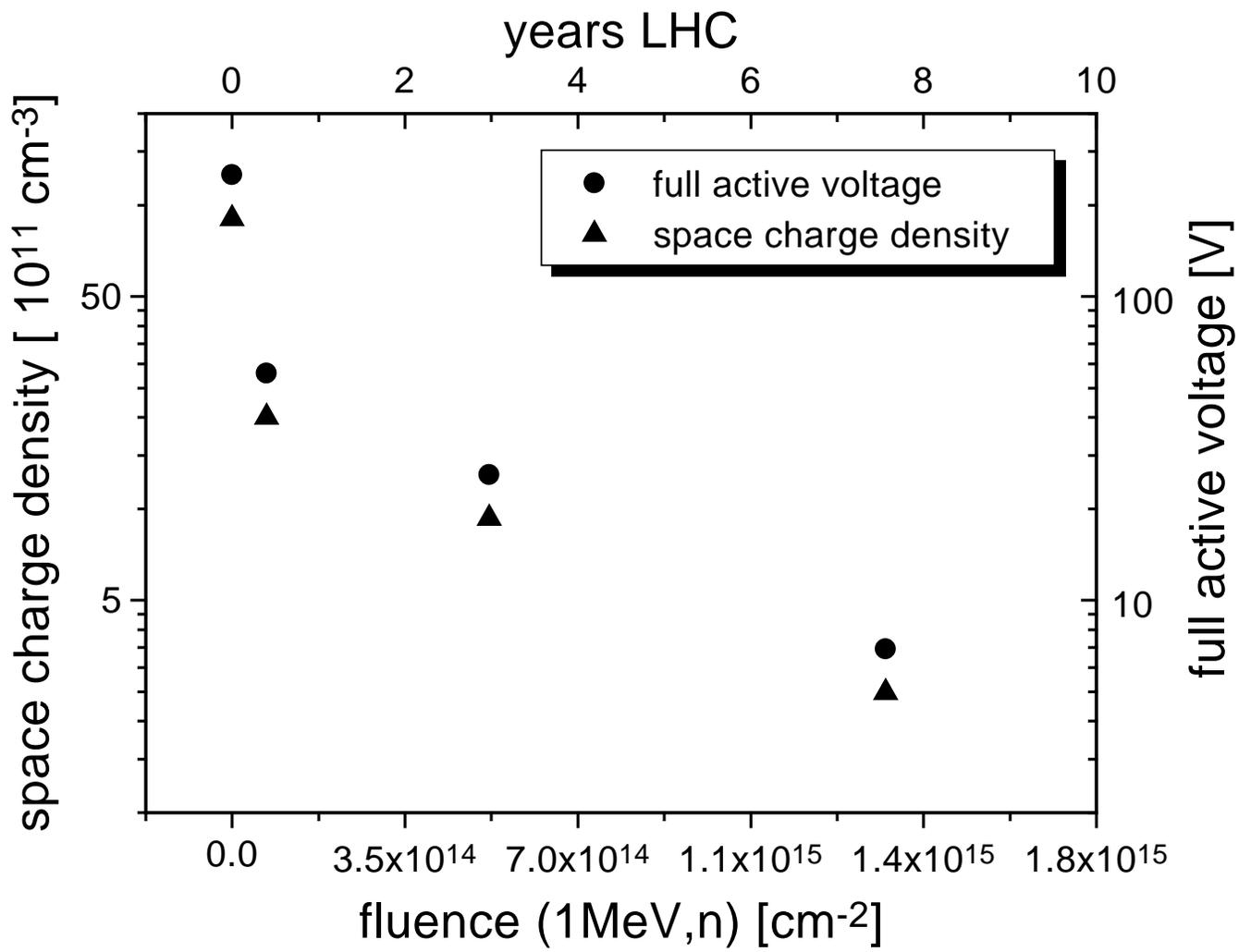}
           }
   \end{center}
\caption{
The space charge density and full active voltage obtained by
CV-measurements versus the irradiation level.
}
\end{figure}

\end{document}